\documentclass[prl,twocolumn,amsmath,amssymb,floatfix,showpacs]{revtex4}
\usepackage{hyperref}
\usepackage{graphicx} 
\usepackage{amsmath}
\usepackage[dvips]{epsfig}
\newcommand{\beq}{\begin{equation}}
\newcommand{\eeq}{\end{equation}} 
\newcommand{\beqa}{\begin{eqnarray}}
\newcommand{\eeqa}{\end{eqnarray}}
\newcommand{\ba}{\begin{array}}
\newcommand{\ea}{\end{array}}

\begin{document}

\title{The onset of nanoscale dissipation in superfluid $^4$He at 
zero temperature: The role of vortex shedding and cavitation}

\author{Francesco Ancilotto$^1$, Manuel Barranco$^2$, Mart\'{\i} Pi$^2$, and Jussi Eloranta$^3$}

\affiliation{$^1$Dipartimento di Fisica e Astronomia ``Galileo Galilei''
and CNISM, Universit\`a di Padova, via Marzolo 8, 35122 Padova, Italy and 
CNR-IOM Democritos, via Bonomea, 265 - 34136 Trieste, Italy\\
$^2$Departament FQA, Facultat de F\'{\i}sica, Universitat de Barcelona. Diagonal 645,
08028 Barcelona, Spain and
Institute of Nanoscience and Nanotechnology (IN2UB),
Universitat de Barcelona.\\
$^3$Department of Chemistry and Biochemistry, 
California State University at Northridge, California 91330, USA }

\begin{abstract} 
Two-dimensional flow past an infinitely long cylinder of nanoscopic radius in
superfluid $^4$He at zero temperature  is studied by time-dependent density
functional theory. The calculations reveal two distinct critical phenomena for
the onset of dissipation: 1) vortex-antivortex pair shedding from the periphery
of the moving cylinder and 2) appearance of cavitation in the wake, which
possesses similar geometry as observed experimentally for fast moving
micrometer-scale particles in superfluid $^4$He. 
Vortex  pairs with the same circulation are
occasionally emitted in the form of dimers, which constitute the building blocks
for the Benard-von Karman vortex street structure observed in classical turbulent
fluids and Bose-Einstein condensates (BEC). 
The cavitation induced dissipation mechanism should be common to all
superfluids that are self-bound and have a finite surface tension, which
include the recently discovered self-bound droplets in ultracold Bose gases.
\end{abstract} 
\date{\today}


\maketitle

One of the manifestations of $^4$He superfluidity at zero temperature ($T$) is the frictionless liquid flow through capillaries at sufficiently low velocities. 
Based on the well-known Landau criterion, the onset of dissipation is related to the unusual form of the superfluid dispersion 
relation, $\epsilon(p)$, which exhibits roton minimum 
$\epsilon(p_{min})$ at $p_{min}$. The flow should become dissipative 
when the velocity reaches the critical Landau value 
$v_L = \epsilon(p_{min}) / p_{min} = 59\textnormal{ m/s}$ \cite{Wil67}. 
Similarly, an object moving in superfluid $^4$He should experience 
drag only above a certain critical velocity threshold $v_c$.
It is well established experimentally that objects moving 
already at much lower velocities than $v_L$ experience drag 
due to the emission of non-linear excitations in the form of 
quantized vortices; see for example Ref. \cite{Wil67}.  

Multiple interacting vortices in a superfluid can form a well-defined lattice or a more complicated vortex tangle, depending on their geometry 
and circulation.
At high vortex densities, vortex 
reconnection events, which are believed to be responsible for 
the large-scale behavior of quantum turbulence \cite{kerr,lead,ogawa,koplik}, become increasingly important. Quantum turbulence
is associated with the proliferation of quantized vortices \cite{vinen,paoletti}. From the  experimental point of view, vorticity can be created by stirring or   
rotating the superfluid \cite{bec_vort,ketter,Kwo16}. 

\begin{figure}
\epsfig{file=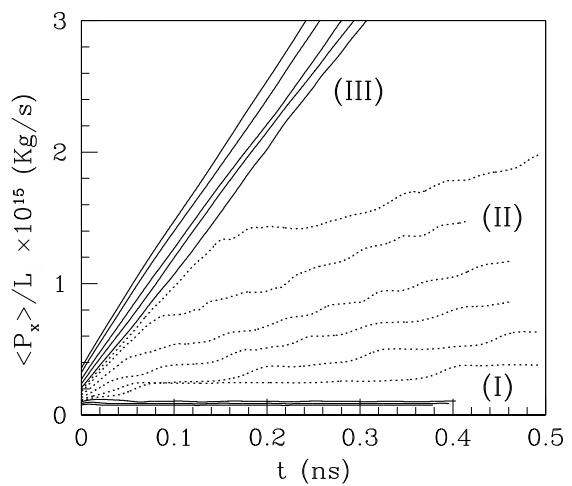,height=2.5 in,clip=}
\caption{
Linear momentum per unit length  around a cylinder with radius $R=3$ \AA{} as a function of time. The curves shown refer to different values of the cylinder velocity. In group (I) they correspond to $v = 0.44, 0.48$, and $0.52$ (in units of $v_L$). In group (II), from bottom to top  they correspond to $v = 0.55, 0.57, 0.61, 0.66, 0.70$, and $0.72$. In group (III), the velocities are from bottom to top $v= 0.74, 0.79, 0.83, 0.89$, and $0.94$.} 
\label{fig1}
\end{figure}

\begin{figure}
\epsfig{file=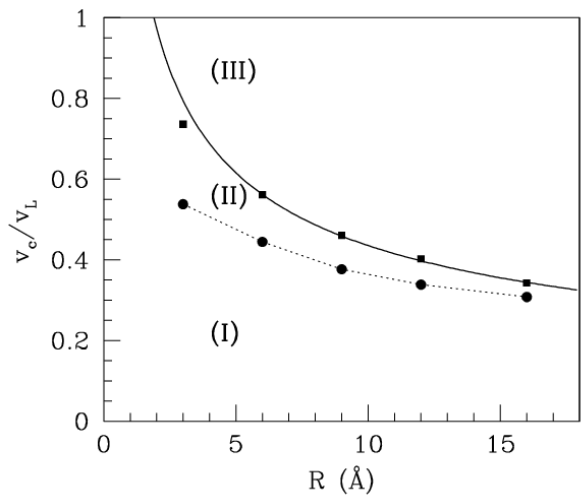,height=2.7 in,clip=}
\caption{
Critical velocities $v_{c1}$ (filled circles) and $v_{c2}$ (filled squares) as a function of the cylinder radius. The solid line shows a fit of $R^{-1/2}$ law to $v_{c2}$ 
whereas the dotted line is just provided as a guide to the eye. The Roman numbers refer to the three different regions identified in Fig. \ref{fig1}.}
\label{fig5}
\end{figure}

\begin{figure}
\epsfig{file=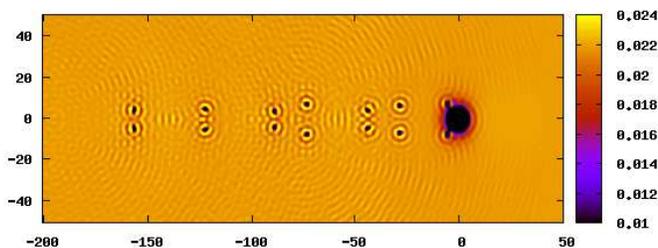,height=1.4 in,clip=}  
\caption{A snapshot of liquid helium density around a moving cylinder ($R = 3$ \AA{}) at constant velocity $v=0.66\,v_L$ (regime II). The cylinder, located at the origin in the $xy$-plane, 
is about to emit a vortex pair.
The lengths are expressed in \AA{} and density contours in \AA$^{-3}$.
}
\label{fig2}
\end{figure}

Although quantized vorticity plays a key role in the onset of dissipation in superfluid flows, a fundamental understanding of their role in exerting drag on 
moving objects and the dependence of the associated critical velocity 
on the object size is still lacking. In this paper, we identify a new, 
previously overlooked, energy dissipation mechanism that takes 
place also well below $v_L$. The energy loss and the induced drag 
force on the object in this mechanism originate from the formation 
of cavitation bubbles in the wake. Cavitation bubbles play a 
crucial role in the appearance of drag as they may act as 
vortex nucleation seeds through local distortions of their  
surface and, more importantly, their nucleation and growth 
provides a significant source of energy loss. 

We have studied the onset of dissipation in superfluid $^4$He at 
$T=0$ by simulating two-dimensional flow past an infinitely 
long cylinder (wire) with a nanoscopic cross-section using 
time-dependent density functional theory (TDDFT). The employed 2D 
wire geometry is not only simpler to simulate than the 
full three-dimensional case associated with, e.g., a 
moving sphere, but it is also appealing because vibrating wire 
resonators are commonly used to study quantum turbulence in 
superfluid $^4$He \cite{Got08,Brad09}. 

The cylinder is represented in the calculations by a 
repulsive external potential, $V(r)=V_0 [1+e^{(r-R)/\sigma}]^{-1}$ 
with $r=\sqrt{x^2+y^2}$, $\sigma = 0.3$ \AA{}, and $V_0=5000$ K. 
This potential represents a hard cylindrical object with 
radius $R$ aligned along the $z$-axis. In the following, 
the liquid flow (or the cylinder motion) is oriented along 
the $x$-axis with a given velocity $v$.

Within DFT, superfluid $^4$He is described by a complex valued 
order parameter (effective wave function) $\Psi( \mathbf{r},t)$, 
which is related to the atomic density 
as $\rho (\mathbf{r},t)= |\Psi( \mathbf{r},t)|^2$. In the 
cylinder frame of reference, the TDDFT equation becomes  
\begin{equation}
\imath \hbar\frac{\partial}{\partial t}  \Psi({\mathbf r},t)  = \left\{
-\frac{\hbar^2}{2m}\nabla^2 + \frac{\delta{\cal
E}_{c}}{\delta\rho}
+ V(r) -v\hat {P}_x  \right\} \,\Psi(\textbf{r},t) 
\label{eq2}
\end{equation}
where $\hat {P}_x = -i\hbar \partial / \partial x $ is the linear 
momentum operator along the $x$-axis and the functional 
${\cal E}_c\left[\rho\right]$ was taken from Ref. \cite{Anc05}. 
This functional includes both finite-range and non-local corrections 
that are required to describe the $T=0$ response of 
liquid $^4$He on the \AA{}ngstr\"om-scale accurately. Due to the translational invariance, 
the problem reduces to finding the fluid density and velocity 
field in the $(x,y)$-plane. The details on solving the TDDFT 
equation can be found from Ref. \cite{Mat11}. 
Previously, DFT models with various levels of complexity have been used to study the motion 
of electrons and ions in liquid $^4$He as well as the mechanism of vortex ring emission; see, e.g., 
Refs. \cite{Ber00,Jin10,Anc10} and references therein.

 
Equation (\ref{eq2}) was solved for wire radii $R=3,6,9,12$, and $16$ \AA{} at 
several fixed values of velocity. The force per unit length  exerted on 
the superfluid by the moving object can be calculated from the 
momentum transfer rate to the fluid
\begin{equation}
F_d = {1\over L} 
{\partial \langle \hat{P}_x\rangle
\over \partial t}
={1\over L} 
{\partial \over \partial t}\left[ \int   d\mathbf{r} \, \Psi^* (\mathbf{r},t)
\hat{P}_x \Psi (\mathbf{r},t)   \right]
\label{eq4}
\end{equation}
where $L$ is the length of the wire. The onset of drag was identified by observing the time dependence of 
$\langle \hat{P}_x \rangle/L$. 

Figure \ref{fig1} shows the time dependence of 
$\langle \hat{P}_x \rangle/L$
for $R=3$ \AA{} at selected values of flow velocity. 
According to Eq. (\ref{eq4}), the drag force acting on the moving object corresponds to
 the slope of the curves shown.
At variance with previously reported results for objects moving 
in superfluid BECs, where a single
 critical velocity separates inviscid flow from the onset 
of drag due to vortex shedding \cite{Kwo16,Kwo15},
we 
find \textit{two distinct critical velocities} for 
superfluid $^4$He. These velocities, which in the following
are denoted by $v_{c1}$ and $v_{c2}$, separate three 
different regimes: (I) inviscid flow, (II) vortex pair shedding, 
and (III) cavitation bubble formation. 
 
For object velocities $v<v_{c1}$ (regime I), 
the fluid profile around 
the object converges rapidly with time into a stationary 
configuration. Both the density and the velocity field for 
such configurations 
are fore-to-aft symmetric. This implies that the drag force 
on the object is zero (flat portion of the curves (I) 
in Fig. \ref{fig1}), which is the well-known D'Alembert 
paradox for classical fluids. Fig. \ref{fig1} also shows 
that in the other two regimes, the time-dependent behavior 
of the total momentum is characterized by notable transient 
periods where the wire experiences a higher drag due to the 
tendency of developing a cavity around the object. 
Furthermore, depending on the wire velocity, the slope may 
then either decrease and settle to a lower value (regime II), or 
remain approximately constant (regime III).

When the velocity exceeds $v_{c1}$, singly-quantized linear 
vortex-antivortex pairs 
are nucleated periodically 
on both sides of the wire cross-section.
The vortices 
eventually detach from the object and drift downstream
as vortex dipoles.
Their appearance is accompanied by drag force 
acting on the moving wire, which increases with 
velocity (group of curves labeled (II) in 
Fig. \ref{fig1}). The oscillations in the dotted curves after
 the initial period reflect the quasi-periodic emission of 
vortex pairs. 
In this regime, the cavity around the object 
largely recovers its circular geometry after each vortex 
emission event. As the velocity is increased, the vortex 
shedding frequency is observed to increase accordingly. 
 Overall, this behavior is similar to BECs 
where quasi-periodic 
vortex-antivortex pair
 emission events also take place \cite{Fri92,Win00,Sch17}.

\begin{figure}
\epsfig{file=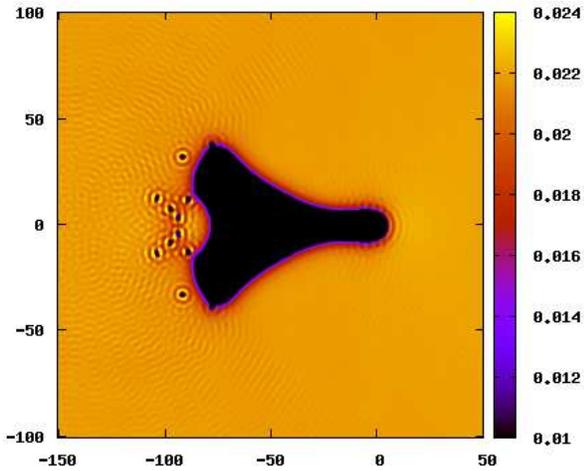,height=2.7 in,clip=}  
\caption{A snapshot of the liquid density around a cylinder ($R = 3$ \AA{}) moving at $v=0.79\,v_L$ (regime III). The cylinder is located at the origin in the $xy$-plane. 
The lengths are expressed in \AA{} and density contours in \AA$^{-3}$.
}
\label{fig3}
\end{figure}

\begin{figure}
\epsfig{file=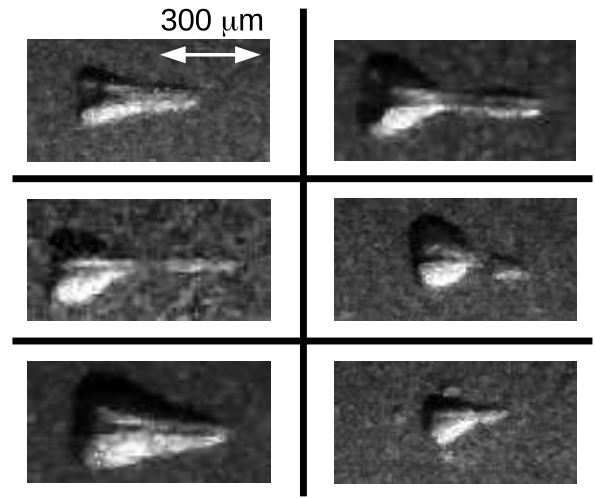,height=2.7 in,clip=}  
\caption{An overview of the bubble shapes observed around fast moving metal particles (diameter few microns) propagating from left to right in 
superfluid $^4$He at 1.7 K (saturated vapor pressure). The data shown correspond to the observations made during the experiments described in Ref. \cite{Bue16}.}
\label{fig7}
\end{figure}

\begin{figure}
\epsfig{file=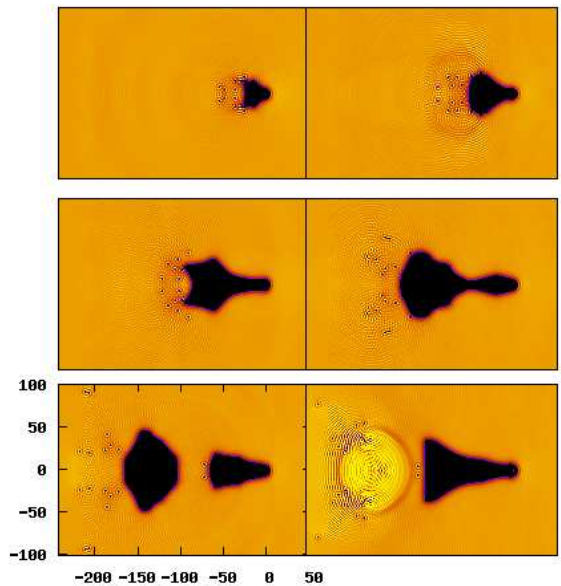,height=3.1 in,clip=}  
\caption{Time evolution of the liquid density around a moving cylinder 
($R=3$ \AA{}) 
at constant velocity $v=0.74\,v_L$ (regime III).
The snapshots are taken between $t=0.15$ ns (top left panel)
and $t=0.65$ ns (bottom right panel).
The lengths are expressed in \AA{} and density contours in \AA$^{-3}$ according to the scale specified in Fig. \ref{fig3}.
} 
\label{fig4}
\end{figure}

Finally, in regime (III) ($v>v_{c2}$), the response of the fluid 
environment close to the moving wire changes dramatically as 
empty cavities are formed in the wake. This cavity formation 
is accompanied by simultaneous quasi-periodic emission of 
vortex pairs.
Note that the transition between 
regimes (II) and (III) is characterized by a discontinuous 
jump in the drag force exerted on the cylinder due to the 
formation of bubbles (see Fig. \ref{fig1}).

Cavitation is initiated by an asymmetric fore-to-aft density profile 
such that the liquid density decreases substantially 
behind the moving wire, thus resulting in a reduction of the 
local pressure. If the velocity is sufficiently high, this 
density decrease can trigger the formation of bubbles 
around or behind the wire. A similar mechanism is also 
responsible for cavitation  in liquid $^4$He at negative pressures \cite{Xio89,Bal02}.


The critical velocities $v_{c1}$ and $v_{c2}$, where the transitions to 
the dissipative regimes (II) and (III) take place, are shown 
in Fig. \ref{fig5} as a function of $R$. At the nanoscale, the velocities 
exhibit distinct dependencies on 
the cylinder radius, but for large radii  
$v_{c1}$ and $v_{c2}$ seem to converge towards a single 
value. Therefore, at the mesoscale, the onset of vortex 
shedding and bubble cavitation should
appear simultaneously at a common critical velocity value $v_c$.
Furthermore, $v_c$ decreases with increasing $R$, which is 
in accordance with the existing experimental data showing 
that the actual critical velocities are lower than
$v_L$
(e.g., cm/s or even mm/s values are 
usually measured) \cite{Wil67,Brad09}.



Figure \ref{fig2} shows a snapshot taken during  
real-time evolution of the system with $R=3$ \AA{} 
and $v = 0.3\,v_L>v_{c1}$ (regime II) \cite{vL}. 
Singly 
quantized vortex pairs with opposite circulation 
(vortex dipoles) 
are emitted behind the moving wire. Note 
that for a spherical object in 3D, the emitted 
vortex dipoles
would be vortex rings instead \cite{Ber00,Jin10,Anc10}. 
The overall vortex emission 
process is quasi-periodic, and the frequency of vortex 
shedding events increases with velocity $v$, which is 
consistent with experimental observations in superfluid $^4$He \cite{Sch17}.

For a cylinder with $R = 3$ \AA{}, Fig. \ref{fig3} shows a 
transient image corresponding to $v = 0.79\,v_L>v_{c2}$ (regime III). 
In addition to vortex dipole emission, formation of a wide 
dynamically evolving cavity is observed. 
Remarkably, in spite of the fact that the shapes and volumes of
such cavities continuously change in time, their 
hydrodynamic drag remains approximately constant.
This is evidenced by the 
fact that the slopes of the solid lines 
in Fig. \ref{fig1} (regime III) do not vary 
significantly as a function of time.

Very similar cavity shapes that appear in regime III have been 
observed in recent experiments where micrometer-scale metal 
particles were injected into bulk superfluid helium by 
laser ablation \cite{Bue16}. The main features of the 
observed fast propagating particle-bubble systems were the 
elongated cavity geometry and the widened flat or 
cone-shaped tail structure (see Fig. \ref{fig7}), which 
are both clearly reproduced by the present simulations. 
Note that an exact match between the 
cavity geometries in the experiments and the present simulations 
is not expected due to the different object geometries 
(i.e., sphere vs. cylinder), the presence of gaseous 
insulating layer between the particle and the 
liquid (i.e., insulating Leidenfrost layer \cite{Vak11}), 
and the difference in length scales (nm vs. $\mu$m). 
Regarding the latter point, Fig. \ref{fig5} shows 
that both vortex emission and cavitation processes in 
these experiments should take place at the same 
value of $v_c$. 

In addition to the velocity dependent cavity shapes formed 
around the particles, a trail of slowly drifting 
cavitation bubbles were also observed behind the fast 
particles propagating in the
 liquid \cite{Bue16}. This is a direct consequence of 
particle-cavitation bubble splitting, which is reproduced 
by the present simulations as demonstrated in Fig. \ref{fig4}. 
When the bubble detaches, the leading portion hosts 
the object and the trailing bubble is left empty. 
As shown by the two bottom panels of Fig. \ref{fig4}, 
the trailing bubble eventually collapses and 
leads to the emission of shock waves.


The moving object may also emit closely spaced vortex 
dimers in the wake, i.e., bound pairs of vortices with 
the same direction of circulation, which are different from 
the vortex dipoles discussed above. 
Once formed, each dimer 
structure remains bound and rotates around its center 
of mass while moving away from the cavity that hosts the wire. 

The formation of vortex dimers
behind the object is the hallmark of well-known 
phenomenon in classical fluid dynamics at 
large Reynolds numbers,
i.e. the Benard-Von Karman (BvK) vortex street, where 
an incompressible flow past an object produces an 
asymmetric wake downstream consisting of quasi-periodically 
nucleated vortex dimers. Such BvK vortex street structures 
have been observed experimentally
in BECs and simulated by the time-dependent Gross-Pitaevskii (GP) 
equation \cite{Kwo16,Sta15}, 
but so far were never observed in superfluid helium. 

The upper critical velocity for 
cavitation, $v_{c2}$, appears to follow the $v_c\propto R^{-1/2}$ 
law as indicated by the solid line in Fig. \ref{fig5}. This behavior 
is similar to the onset of drag when liquid $^4$He is forced to 
flow through a cylindrical channel of diameter $d$ where the 
critical velocity was found to scale as $v_c\sim d^{-1/2}$ \cite{Bis77}. 
In BECs, the experiments of 
Kwon \textit{et al.} \cite{Kwo15}, where vortex
shedding was produced by a laser beam of ``radius" $R$ moving 
through the condensate, show a $1/R^s$ dependence of the critical velocity with $s < 1$.

Several studies have, however, suggested a scaling law
$v_c\propto 1/R$. 
For example, the superfluid Reynolds number $\textnormal{Re}_s$ 
was introduced \cite{scho16,reeves,finne} by
replacing the kinematic viscosity $\nu $ with quantized circulation $\kappa = h/m$
in the definition of Reynolds number, $\textnormal{Re}=Dv/\nu$, 
yielding $\textnormal{Re}_s \sim mvD/h$ where $D$ is the characteristic 
size of the system. This model has been employed to analyze 
oscillating sphere data in superfluid $^4$He in the mK 
regime \cite{scho16,reeves,finne} where
a critical $\textnormal{Re}_s$ value for the appearance of turbulent behavior was determined.
Similar conclusions were also drawn from simulations for 
the onset of turbulent flow in BECs \cite{reeves,zwerger}.
The existence of a threshold value for $\textnormal{Re}_s$ implies 
that the associated critical velocity $v_c$ must scale as $1/R$,
which is different from our result for superfluid helium.
At present, the situation regarding the scaling law thus remains inconclusive.



In summary, we have shown that
the formation of cavitation bubbles plays an important role in the onset of dissipation 
 in superfluid $^4$He, 
which is at variance with the accepted view that
only vorticity should be responsible for such behavior.
%
It is worth noticing that the reported cavitation 
dissipation mechanism is not applicable to cold gas BECs 
because they are not self-bound and have no surface tension. 
However, the recently observed self-bound droplets in ultracold dipolar 
bosonic gas \cite{Sch16}
could provide an interesting test ground for exploring this mechanism further.

\begin{acknowledgments}
We thank Antonio Mu\~noz and Leticia Tarruell 
for useful discussions. This work was 
performed under Grants No 2014SGR401 from Generalitat de Catalunya, 
FIS2014-52285-C2-1-P from DGI (Spain) and DMR-1205734 from NSF (USA).
\end{acknowledgments}


\begin{thebibliography}{99}
\bibitem{vL}
The Landau critical velocity for the applied functional (see Ref. \cite{Anc05}) is $v_L = 94.4$ m/s, which is larger that the experimental value (59 m/s). 
Despite this difference, our simulations show that neither vortex nor bubble nucleation are connected to the presence of roton minimum or the associated 
Landau velocity. To facilitate comparison with experiments, we present the results relative to $v_L$ of the functional.
\end{thebibliography}
\end{document}